\documentclass[5p]{elsarticle}
\usepackage{amsmath}
\usepackage{amssymb}
\usepackage{graphicx}

\newcommand{\ve}{\varepsilon}

\begin{document}

\title{Cross-Section Fluctuations in Chaotic Scattering}

\author{B.~Dietz\fnref{ikp}}

\author{H.~L.~Harney\fnref{mpi}}

\author{A.~Richter\corref{cor}\fnref{ikp,ect}}
\ead{richter@ikp.tu-darmstadt.de}

\author{F.~Sch{\"a}fer\fnref{ikp}}

\author{H.~A.~Weidenm{\"u}ller\fnref{mpi}}

\address[ikp]{Institut f{\"u}r Kernphysik, Technische Universit{\"a}t
Darmstadt, D-64289 Darmstadt, Germany}
\address[mpi]{Max-Planck-Institut f{\"u}r Kernphysik, D-69029 Heidelberg,
Germany}
\address[ect]{$\rm ECT^*$, Villa Tambosi, I-38100 Villazzano (Trento), Italy}

\cortext[cor]{Corresponding author}

\date{\today}

\begin{abstract}

For the theoretical prediction of cross--section fluctuations in
chaotic scattering, the cross-section autocorrelation function is
needed. That function is not known analytically. Using experimental
data and numerical simulations, we show that an analytical
approximation to the cross-section autocorrelation function can be
obtained with the help of expressions first derived by Davis and
Boos\'e. Given the values of the average $S$--matrix elements and the
mean level density of the scattering system, one can then reliably
predict cross-section fluctuations.

\end{abstract}


\maketitle

\section{Purpose}

Quantum chaotic scattering is an ubiquitous phenomenon. It occurs, for
instance, in nuclear physics~\cite{Ver85}, in electron transport
through disordered mesoscopic samples~\cite{Bee97}, and in microwave
billiards~\cite{Die08}. In all cases, the cross section displays
random fluctuations versus energy or frequency. These are due to the
random features of the underlying resonances.  With $d$ the average
resonance spacing and $\Gamma$ the average width, data on
cross-section fluctuations exist for the entire range of the parameter
$\Gamma / d$, from the regime of isolated resonances ($\Gamma \ll d$)
to that of strongly overlapping resonances ($\Gamma \gg d$). The
analysis of the data focuses on the value of the average cross section
and on quantities that characterize the cross-section 
fluctuations. These are the variance of the cross section and certain
correlation functions. For the analysis, one needs
theoretical expressions for these quantities. These should be generic
and only use a minimum of adjustable parameters.

The generic theoretical treatment of chaotic scattering employs a
combination of scattering theory and random-matrix theory~\cite{Ver85}
and uses as input the values of $d$ and of the energy-averaged elements
$\overline{S}$ of the scattering matrix $S$. Analytical results exist
for the $S$-matrix autocorrelation and cross-correlation
functions~\cite{Ver85} (including the value of the average cross
section) and for the third and fourth moments of the
$S$-matrix~\cite{Dav88,Dav89}. Because of the complexity of the problem,
analytical results for higher moments of the cross section or for
cross-section correlation functions cannot be expected in the
foreseeable future.

The present paper aims at filling that gap. We combine the available
analytical information~\cite{Ver85,Dav88,Dav89}, results of computer
simulations, and of experimental work on microwave
billiards~\cite{Die08,Die09} to study the cross-section
autocorrelation function for all values of $\Gamma / d$. In
particular, we address the following questions. (i) For which values
of $\Gamma / d$ and with which accuracy can the cross-section
autocorrelation function be predicted in terms of the $S$-matrix
autocorrelation function? (ii) Which analytical alternatives exist
should that approach fail?

\section{Framework}

We consider chaotic scattering in a time-reversal invariant system
described by a unitary and symmetric scattering matrix $S_{a b}(E)$.
Here $a,b=1,\ldots,\Lambda$ denote the channels and $E$ the energy (or,
in the case of microwave billiards, the frequency). The number
$\Lambda$ of channels may range from unity to a large number, $\Lambda
\gg 1$. Chaotic scattering is modeled by writing the $S$-matrix in the
form~\cite{Mah69}
\begin{equation}
	S_{ab}(E) = \delta_{ab} - i \sum_\mu W_{a\mu}D^{-1}_{\mu\nu}(E)
	W_{b\nu}
	\label{1a}
\end{equation}
where
\begin{equation}
	D_{\mu \nu}(E) = E\,\delta_{\mu \nu} - H_{\mu\nu} + i \pi\sum_c
	W_{c\mu} W_{c\nu} \ .
	\label{1b}
\end{equation}
The real and symmetric Hamiltonian matrix $H$ has dimension $N \gg 1$
and describes the dynamics of $N$ resonances labeled by Greek letters.
These are coupled to the channels by the real matrix elements $W_{a
\mu}$. Chaos is taken into account by choosing $H$ as a member of the
Gaussian orthogonal ensemble (GOE) of random matrices~\cite{Ver85}.
Thus, the elements of $H$ are Gaussian random variables with zero mean
values and second moments given by $\overline{H_{\mu \nu} H_{\rho
\sigma}} = (\lambda^2/N) [\delta_{\mu \rho} \delta_{\nu \sigma} +
\delta_{\mu \sigma} \delta_{\nu \rho}]$. Here and in the sequel, the
overbar denotes the average over the ensemble. The parameter $\lambda$
determines (or is determined by) the average level spacing $d$ of the
$N$ resonances. It is convenient to decompose $S(E)$ into an average
and a fluctuating part,
\begin{equation}
	S_{a b}(E) = \overline{S_{a a}} \ \delta_{a b} + S^{\rm fl}_{a
	b}(E) \ ,
	\label{1}
\end{equation}
where the average $S$-matrix $\overline{S}$ is assumed to be
diagonal. The values of the diagonal elements $\overline{S_{a a}}$
serve as input parameters for the statistical model and are assumed to
be known. That is the typical case: In nuclei, $\overline{S_{a a}}$ is
given in terms of the optical model of elastic scattering, in
microwave billiards $\overline{ S_{a a}}$ is determined by the running
average over a measured spectrum \cite{Die08,Die09}. In rare cases,
the average $S$-matrix may not be diagonal. By an orthogonal
transformation in channel space, $\overline{S}$ can be reduced to
diagonal form, see Refs.~\cite{Dav89,Eng73}. For simplicity we do not
address that case. By the same transformation, the phases of the
$S$-matrix usually appearing as factors on the right--hand side of
Eq.~(\ref{1}), can be removed. Both for the S-matrix model
Eq.~(\ref{1a}) considered in the present work and the experimental data
the average $S$-matrix is real and diagonal. Starting from
Eq.~(\ref{1a}), the $S$-matrix autocorrelation function (or ``two-point
function'')
\begin{equation}
	C^{(2)}_{a b}(\ve) = \overline{ S^{\rm fl}_{a b}(E - \ve/2)
	S^{\rm fl *}_{a b}(E + \ve/2)}
	\label{2}
\end{equation}
has been calculated analytically~\cite{Ver85} for $N \gg 1$ and fixed
$\Lambda$. The resulting expression depends only on the difference
$\ve$ of the two energy arguments, on the average level spacing $d$ of
the system, and on the transmission coefficients $T_a$ of all channels
$a$ defined by
\begin{equation}
	T_a = 1 - |\overline{S_{a a}}|^2 \ .
	\label{3}
\end{equation}
The transmission coefficients obey $0 \leq T_a \leq 1$. These
coefficients measure the unitarity deficit of the average $S$-matrix
and give the probability with which the resonances take part in the
reaction. This is seen by using the decomposition~Eq.~(\ref{1}) and
the definition Eq.~(\ref{3}) to write the unitarity condition for $S$
in the form
\begin{equation}
	T_a = \sum_b |S^{\rm fl}_{a b}(E)|^2 \ .
	\label{3a}
\end{equation}
For $T_a = 0$ or $|\overline{S_{a a}}| = 1$, we have $S^{\rm fl}_{a
b}(E) = 0$ for all $b$, and the resonances are not reached from
channel $a$.  Conversely, $\sum_b |S^{\rm fl}_{a b}(E)|^2$ is maximal
for $T_a = 1$ or $\overline{S_{a a}} = 0$ (complete absorption of the
incident flux in channel $a$ by resonance formation). The transmission
coefficients $T_a$ determine the average width $\Gamma$ of the
resonances. An approximation for $\Gamma$ is the ``Weisskopf estimate''
\begin{equation}
	\Gamma = \frac{d}{2 \pi} \sum_a T_a \ .
	\label{3b}
\end{equation}
The case of strongly overlapping resonances $\Gamma \gg d$ (``Ericson
regime''~\cite{Eri60,Eri63,Bri63}) occurs for $\sum_a T_a \gg 1$: The
number $\Lambda$ of channels must be large and most of the individual
transmission coefficients $T_a$ must not be small. Conversely, the
case of nearly isolated resonances $\Gamma\ll d$ is realized when
$\Lambda$ is of order unity or when $\Lambda$ is large but all $T_a$
are small. The theory developed in Ref.~\cite{Ver85} and used in
Refs.~\cite{Dav88,Dav89} applies to all values of $\Gamma / d$.
Equation~(\ref{3b}) is exact in the Ericson regime and fairly reliable
elsewhere.

Under omission of kinematical factors the cross section in nuclear
physics, the conductance in electron transport and the transmitted
power in microwave billiards are all given by $|S_{a b}(E)|^2$ or by a
sum of such terms. For brevity we refer to $|S_{a b}(E)|^2$ as to the
cross section. The average cross section $\overline{|S_{a b}(E)|^2}=
|\overline{S_{a a}(E)}|^2 \delta_{a b} + \overline{|S^{\rm fl}_{a
b}(E)|^2}$ is given in terms of $\overline{S_{a a}}$ and of
$C^{(2)}_{a b}(0)$ and is, thus, known.  Fluctuations of the cross
section are measured in terms of the cross-section autocorrelation
function
\begin{equation}
	{\cal C}_{a b}(\ve) = \overline{|S_{a b}(E+\ve/2)|^2 |S_{a b}(E -
	\ve/2)|^2} - \overline{|S_{a b}|^2}^2 \ .
	\label{4}
\end{equation}
That function is the object of central interest in the present paper.
With the help of the decomposition~Eq.~(\ref{1}) we write
\begin{align}
	{\cal C}_{a b}(\ve) &= 2\delta_{a b}\mathfrak{Re}\bigg\{\overline{S_{a a}}^2 
	C^{(2)}_{a a}(\ve) \nonumber \\
	&+\overline{S_{a a}} \ \overline{S^{\rm fl *}_{a a}(E+
	\ve/2) |S^{\rm fl}_{a a}(E-\ve/2)|^2} 
	\nonumber \\
	&+\overline{S_{a a}} \ \overline{S^{\rm fl *}_{a a}(E-
	\ve/2) |S^{\rm fl}_{a a}(E+\ve/2)|^2}\bigg\}
	\nonumber \\
	&+\overline{|S^{\rm fl}_{a b}(E+\ve/2)|^2 |S^{\rm fl}_{a b}(E -
	\ve/2)|^2} - \overline{|S^{\rm fl}_{a b}|^2}^2\,.
	\label{5}
\end{align}
We have used that in the experiments and in the considered $S$-matrix model
(Eq.~(\ref{1a})) $\overline{S_{aa}}$ is real, that by
definition $\overline{S^{\rm fl}_{ab}(E)} = 0$ and that $\overline{
S^{\rm fl}_{a b}(E_1) S^{\rm fl}_{a b}(E_2)} = 0$ for all $a, b$ and all
$E_1, E_2$. The last relation holds because all poles of $S$ lie in the
lower half of the complex energy plane. To determine ${\cal C}_{a
b}(\ve)$ we need to know the four-point function
\begin{equation}
	C^{(4)}_{a b}(\ve) = \overline{|S^{\rm fl}_{a b}(E+\ve/2)|^2 |S^{\rm
	fl}_{a b}(E - \ve/2)|^2} - \overline{|S^{\rm fl}_{a b}|^2}^2
	\label{7}
\end{equation}
and, in the elastic case $a = b$, also the three-point function
\begin{equation}
	C^{(3)}_{a b}(\ve) = \overline{S^{\rm fl *}_{a b}(E+\ve/2)
	|S^{\rm fl}_{a b}(E-\ve/2)|^2} \ .
	\label{6}
\end{equation}
These functions are known analytically only for $\ve =0$ and in the
Ericson regime ($\Gamma \gg d$), see below.

To determine magnitude and $\ve$-dependence of ${\cal C}_{a b}(\ve)$, we
combine analytical results with numerical and experimental evidence as
follows. (i) Analytical results: In Refs.~\cite{Dav88,Dav89} analytic
expressions are given for two functions $F^{(4)}_{a b}(\ve)$ and
$F^{(3)}_{a b}(\ve)$ that look similar to but actually differ from
$C^{(4)}_{a b}(\ve)$ and $C^{(3)}_{a b}(\ve)$, respectively. These are
defined by
\begin{align}
	F^{(4)}_{ab}(\ve)&=\overline{\left[S^{\rm fl *}_{a b}(E+\ve/2)\right]^2
	\left[S^{\rm fl}_{a b}(E - \ve/2)\right]^2} \ , \nonumber\\
	F^{(3)}_{a b}(\ve)&=\overline{S^{\rm fl *}_{a b}(E+\ve/2)
	\left[S^{\rm fl }_{a b}(E - \ve/2)\right]^2} \, .
	\label{Feps}
\end{align}
We note that in $C^{(4)}_{a b}(\ve)$ and in $C^{(3)}_{a b}(\ve)$ the
elements $S^{\rm fl}_{a b}$ and $S^{\rm fl *}_{a b}$ carry pairwise the
same energy arguments. This is not the case for $F^{(4)}_{a b}(\ve)$ and
$F^{(3)}_{a b}(\ve)$. However, $F^{(3)}_{ab}(0)$ coincides with
$C^{(3)}_{a b}(0)$, and $F^{(4)}_{ab}(0)$ differs from $C^{(4)}_{a
b}(0)$ only by the known term $\overline{|S^{\rm fl}_{a b}|^2}^2$. We
are going to show that for $\ve\ne 0$ it is possible to approximate
$C^{(4)}_{a b}(\ve)$ in terms of $F^{(4)}_{a b}(\ve)$, and under certain
conditions $C^{(3)}_{a b}(\ve)$ in terms of $F^{(3)}_{a b}(\ve)$. For
the convenience of the reader we, therefore, give in the Appendix
analytic expressions for $F^{(n)}_{a b}(\ve)$ for $n = 2, 3, 4$, where
$F^{(2)}_{a b}(\ve)\equiv C^{(2)}_{a b}(\ve)$. We briefly show how the
threefold integrals can be evaluated numerically to avoid the apparent
singularities of the integrand. (ii) Numerical results: For the
numerical simulations we use Eqs.~(\ref{1a}) and (\ref{1b}) and fixed
values for $\lambda$, for the transmission coefficients $T_a$, and for
$N$ as initial values. Calculations were typically done for several 100
realizations to minimize statistical errors. The results agree very well
with the available analytical results but go beyond them. (iii) Data:
The data stem from measurements of transmission and reflection
amplitudes of microwaves in a flat cylindrical resonator made of copper
and coupled to two antennas, see Refs.~\cite{Die08,Die09}. Microwave
power was coupled into the resonator with the help of a vector network
analyzer. The range of the excitation frequency was chosen such that
only one vertical electric field mode is excited.  Then the microwave
cavity simulates a two-dimensional quantum
billiard~\cite{Stoeckmann1990,Richter92}. The resonator had the shape of a tilted
stadium billiard whose classical dynamics is chaotic. Transmission and
reflection amplitudes correspond to complex $S$-matrix elements that are
theoretically modeled by Eqs.~(\ref{1a}) and (\ref{1b})
\cite{Die08,Die09}.

\section{Inelastic case ($a \neq b$)}
\label{inel}

The inelastic case is simpler than the elastic one because it involves
only the function $C^{(4)}_{a b}(\ve)$, see Eq.~(\ref{5}). We begin
with the Ericson regime $\Gamma \gg d$, see
Refs.~\cite{Eri60,Eri63,Bri63}. In Ref.~\cite{Bri63} it was suggested
and in Ref.~\cite{Aga75} it was shown that for $\Gamma \gg d$ the
fluctuating $S$-matrix elements $S^{\rm fl}_{a b}$ possess a bivariate
Gaussian distribution centered at zero. That fact implies that all
higher moments and correlation functions can be computed from
$C^{(2)}_{a b}(\ve)$ by way of Wick contraction. Hence $C^{(3)}_{a
b}(\ve) = 0$ and
\begin{equation}
	C^{(4)}_{a b}(\ve) = |C^{(2)}_{a b}(\ve)|^2 \ .
	\label{9}
\end{equation}
In the Ericson regime, the two-point function has the
value~\cite{Eri60,Aga75}
\begin{equation}
	C^{(2)}_{a b}(\ve) = (1 + \delta_{a b}) \ \frac{T_a T_b}
        {\sum_c T_c + 2 i \pi \ve / d} \ .
	\label{8}
\end{equation}
Thus, for $a \neq b$ and $\Gamma \gg d$, the cross-section
autocorrelation function is known analytically. It has the shape of a
Lorentzian with width $\Gamma$ as given by Eq.~(\ref{3b}).

\begin{figure}[ht]
	\includegraphics[width=8.5cm]{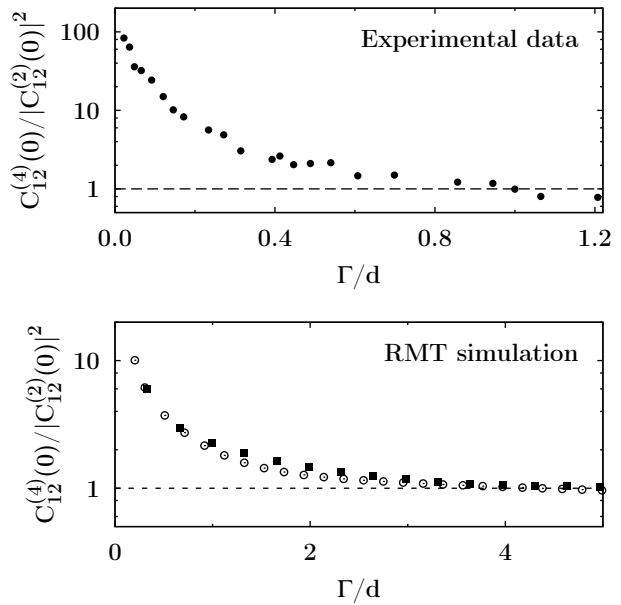}
	\caption{Ratios of four-point- and squared two-point-functions
	of $S^{\rm fl}_{12}$ on a semi-logarithmic plot for three
	cases. Upper panel: Experimental data~\cite{Die08,Die09} (filled
	circles). Lower panel: Data generated numerically from Eq.~(\ref{1a})
	for $\Lambda =32$ channels (open circles) and for $\Lambda =52$
	channels (full squares) with identical transmission coefficients.}
	\label{fig1}
\end{figure}

How far can we use Eq.~(\ref{9}) outside the Ericson regime, i.e., for
smaller values of $\Gamma / d$? Figure~\ref{fig1} shows the ratio
$C^{(4)}_{1 2}(0) / |C^{(2)}_{1 2}(0)|^2$ versus $\Gamma / d$ as a
function of $\Gamma / d$ on a semi--logarithmic plot for three cases as
indicated in the figure caption. Case (i) with unequal transmission
coefficients is obtained from experimental data and suggests that
Eq.~(\ref{9}) holds approximately for $\Gamma \gtrsim d$. The two other
cases result from numerical simulations with, respectively, $\Lambda
=32$ and $\Lambda=52$ equal transmission coefficients and imply, that
Eq.~(\ref{9}) holds for $\Gamma \gtrsim 3d$. The ratio $C^{(4)}_{1 2}(0)
/ |C^{(2)}_{1 2}(0)|^2$ increases dramatically with decreasing $\Gamma /
d$ so that Eq.~(\ref{9}) cannot be used much below $\Gamma \approx d$.

The origin of the failure of Eq.~(\ref{9}) for small values of $\Gamma /
d$ is easily understood. Qualitatively speaking, it is intuitively clear
that cross-section fluctuations (measured in units of the average cross
section) are much larger for isolated than for overlapping resonances.
Quantitatively, the assumption that underlies Eq.~(\ref{9}) is that the
distribution of the fluctuating $S$-matrix elements $S^{\rm fl}_{a
b}(E)$ is Gaussian. That assumption holds only if the width $\gamma$ of
the distribution is sufficiently small. Indeed, the unitarity condition
Eq.~(\ref{3a}) implies that for all $a$ and $b$ we must have $|S^{\rm
fl}_{a b}|^2 \leq T_a \leq 1$, and $\gamma$ must be so small that the
contribution of the tails of the distribution that extend beyond the
values $\pm T^{1/2}_a$, is negligible. Otherwise, significant deviations
from a Gaussian distribution are to be expected. To estimate $\gamma$ in
the Ericson regime we use Eq.~(\ref{8}) which for $a \neq b$ gives
$\gamma = |C^{(2)}_{a b}(0)|^{1/2} = [T_a T_b / \sum_c T_c]^{1/2}$.
Thus, since $\sum_c T_c \gg 1$ this estimate yields $\gamma \ll 1$ as
expected. For smaller values of $\Gamma / d$, constraints on the
distribution of $S^{\rm fl}_{a b}$ due to the unitarity
condition~(\ref{3a}) are expected to become increasingly important as
the number $\Lambda$ of terms in the sum in Eq.~(\ref{3a}) decreases.

\begin{figure}[ht]
	\includegraphics[width=8.5cm]{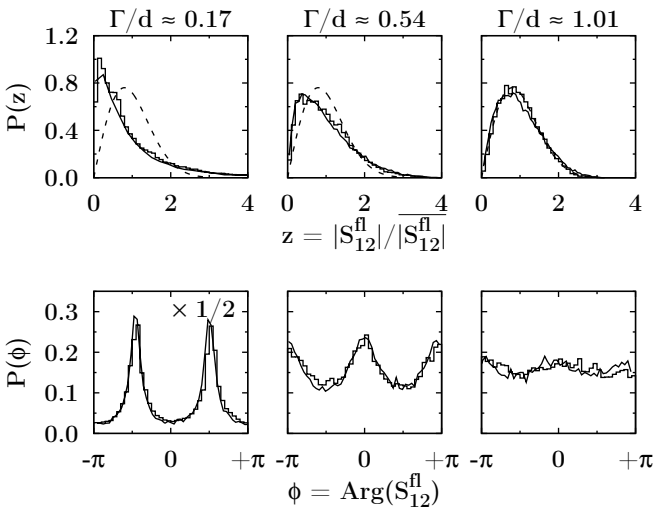}
	\caption{Distribution of the modulus (upper row) and phase (lower row)
	of $S^{\rm fl}_{12}$ obtained from the experimental data (histograms)
	and from numerical simulations (solid lines) for three values of
	$\Gamma / d$. In the upper row, the bivariate Gaussian distribution
	expected in the Ericson limit is shown as a dashed line. The
	transmission coefficients have the following values~\cite{Die08,Die09}.
	For $\Gamma/d=0.17$: $T_1=0.097,\, T_2=0.130,\, \tau_{\rm abs}=0.810$,
	for $\Gamma/d=0.54$: $T_1=0.417,\, T_2=0.475, \tau_{\rm abs}=2.274$,
	for $\Gamma/d=1.01$: $T_1=0.784,\, T_2=0.665, \tau_{\rm abs}=4.903$.}
	\label{fig2a}
\end{figure}

\begin{figure}[ht]
	\includegraphics[width=8.5cm]{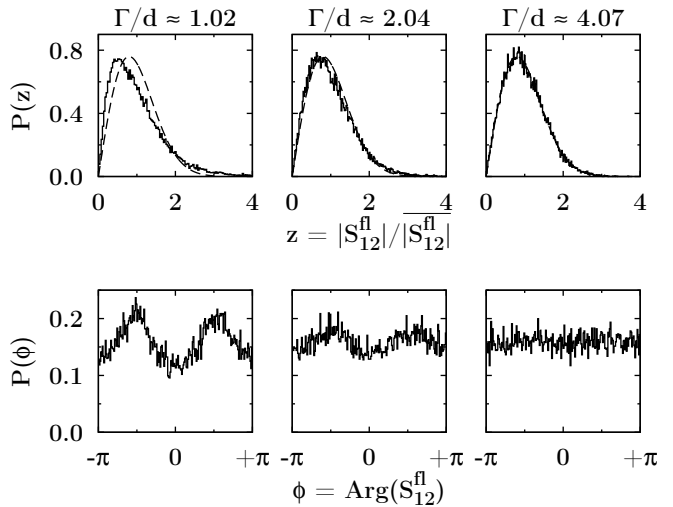}
	\caption{Distribution of the modulus (upper row) and phase (lower row)
	of $S^{\rm fl}_{12}$ obtained from numerical simulations (histograms)
	for $\Lambda = 32$ channels with identical transmission coefficients
	for three values of $\Gamma / d$. In the upper row, the bivariate
	Gaussian distribution expected in the Ericson limit is shown as a
	dashed line.}
	\label{fig2b}
\end{figure}

We have checked this explanation by investigating the distribution of
$S^{\rm fl}_{1 2}$. A bivariate Gaussian distribution for $S^{\rm fl}_{1
2}$ implies that the distribution of $z = |S^{\rm fl}_{1 2}| /
\overline{|S^{\rm fl}_{1 2}|}$ has the form $P(z) = (\pi/2) z
\exp[-(\pi/4)z^2]$, and that the phase of $S^{\rm fl}_{1 2}$ is
distributed uniformly in the interval $\{ 0, 2 \pi\}$. In
Fig.~\ref{fig2a} and Fig.~\ref{fig2b} and for values of $\Gamma / d$
indicated above each panel, we compare in the upper row the function
$P(z)$ with experimental and numerical data, respectively. In the panels
in the lower row, we show the corresponding distributions of the phase.
The numerical data in the three panels of Fig.~\ref{fig2a} are obtained
by simulating absorption in the resonator in terms of a large number of
fictitious channels with small transmission coefficients in each
channel. Then their sum $\tau$ is the only parameter. It was determined
as described in \cite{Die08} from a fit of the experimental
autocorrelation function to the analytic result given in \cite{Ver85}.
The data show that agreement with the bivariate Gaussian distribution is
attained for $\Gamma / d \gtrsim 1$. The data in Fig.~\ref{fig2b}
(generated numerically for 32 equal channels) show agreement with the
bivariate Gaussian distribution only for $\Gamma / d \gtrsim 3$. This
suggests that the limit of a bivariate Gaussian distribution for $S^{\rm
fl}_{a b}$ with $a\ne b$ is attained for larger values of $\Gamma /d$
when all transmission coefficients are equal than when the transmission
coefficients differ. These results account for the deviations from
Eq.~(\ref{9}) displayed in Fig.~\ref{fig1}.


\begin{figure}[ht]
	\includegraphics[width=8.5cm]{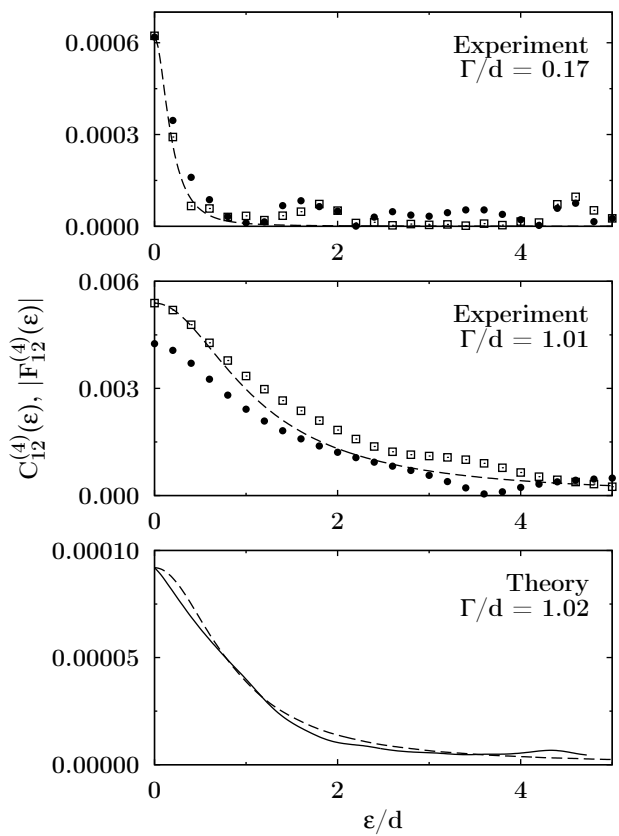}
	\caption{The dependence on $\ve$ of $C^{(4)}_{1 2}(\ve)$ and of the
	renormalized function $|F^{(4)}_{1 2}(\ve)|$ for three cases with
	$\Gamma / d$ as indicated in the panels. In the upper two panels we
	show experimental curves for $C^{(4)}_{1 2}(\ve)$ (filled circles) and
	for $|F^{(4)}_{1 2}(\ve)|$ (open squares), in the lowest panel a
	numerically generated curve for $C^{(4)}_{1 2}(\ve)$ (solid line) for
	32 identical channels. In all three panels, the analytic function
	$|F^{(4)}_{1 2}(\ve)|$ is shown as dashed line. Both the experimental and
	the analytic values for $|F^{(4)}_{1 2}(\ve)|$ are renormalized with a
	factor determined from the analytic value for $C^{(4)}_{1 2}(0)$.}
	\label{fig3}
\end{figure}

Concerning the $\ve$-dependence of $C^{(4)}_{1 2}(\ve)$ and of
$|C^{(2)}_{1 2}(\ve)|^2$, our data show that for those values of $\Gamma
/ d$ where the ratio $C^{(4)}_{1 2}(0) / |C^{(2)}_{1 2}(0)|^2 \approx
1$, that dependence is sufficiently similar so that $C^{(4)}_{1 2}(\ve)$
can be reliably approximated by $|C^{(2)}_{1 2}(\ve)|^2$. (By that we
mean that the full two--point function defined in Eq.~(\ref{2}) and not
the approximate form Eq.~(\ref{8}) has to be used). That leaves us with
the question how to approximate $C^{(4)}_{1 2}(\ve)$ analytically for
$\Gamma \lesssim d$. An obvious possibility is offered by the function
$F^{(4)}_{1 2}(\ve)$ defined in the first of Eqs.~(\ref{Feps}).
Figure~\ref{fig3} shows experimental (upper two panels) and numerically
(lowest panel) generated values of $|F^{(4)}_{1 2}(\ve)|$ and of
$C^{(4)}_{1 2}(\ve)$ versus $\ve$ for two different values of $\Gamma /
d$, and  for $\Lambda = 32$ channels with identical transmission
coefficients $T_a=0.2$ and a value of $\Gamma / d$ close to the largest
achieved in the experiment, respectively. The function $|F^{(4)}_{1
2}(\ve)|$ was rescaled so that at $\ve = 0$ it agrees with $C^{(4)}_{1
2}(0)$. In the upper two panels the experimental curves for $C^{(4)}_{1
2}(\ve)$ (filled circles) and the rescaled function $|F^{(4)}_{1
2}(\ve)|$ are shown together with the analytic result for $|F^{(4)}_{1
2}(\ve)|$ multiplied with the same scaling factor as the experimental
one. In the lowest panel we compare the numerical result for $C^{(4)}_{1
2}(\ve)$ with the analytic rescaled result for $|F^{(4)}_{1 2}(\ve)|$.
We note that the agreement is excellent for all three cases. Similarly
good agreement was found also for $\Lambda = 52$ channels and several
values of $\Gamma / d$. We conclude that we have reached a simple and
reliable prescription for approximating $C^{(4)}_{1 2}(\ve)$
analytically for all values of $\Gamma / d$: Calculate $C^{(4)}_{1
2}(0)$ and $|F^{(4)}_{1 2}(\ve)|$ analytically. (The formulas
needed~\cite{Dav88,Dav89} are given in the Appendix). Rescale
$|F^{(4)}_{1 2}(\ve)|$ so that its value at $\ve = 0$ agrees with
$C^{(4)}_{1 2}(0)$. The resulting rescaled function $|F^{(4)}_{1
2}(\ve)|$ is an excellent approximation to $C^{(4)}_{1 2}(\ve)$ for all
values of $\Gamma / d$.

\section{Elastic Case ($a = b$)}

The elastic case is analogous to the inelastic one only if
$\overline{S_{a a}} = 0$. Then all conclusions drawn in
Section~\ref{inel} apply. In general, that is not the case. In
particular, for $\Gamma \ll d$ (isolated resonances) the last term in
Eq.~(\ref{1a}) tends to zero and $\overline{S_{a a}}$ approaches unity.
The unitarity constraint on $S$ implies $| \overline{S_{a a}} + S^{\rm
fl}_{a a}| \leq 1$ and the distribution of $S^{\rm fl}_{a a}$ must then
become skewed. Therefore, we have to expect that Eq.~(\ref{9}) applies
less generally in the elastic than in the inelastic case, and that
$C^{(3)}_{a a}(\ve )$ in Eq.~(\ref{5}) plays an important role.
Actually, it was pointed out in Refs.~\cite{Dav88,Dav89} that even in
the Ericson limit, a profound difference between the elastic and the
inelastic cases exists: For $\Gamma \gg d$, $\overline{|S^{\rm fl}_{a
a}|^3}$ and $\overline{ |S^{\rm fl}_{a a}|^4}$ have similar values
unless $T_a \approx 1$ or $\overline{S_{a a}} \approx 0$. That shows
that in order to predict ${\cal C}_{a a}(\ve)$ we need to know both
$C^{(3)}_{a a}(\ve)$ and $C^{(4)}_{a a}(\ve)$ for all values of $\Gamma
/ d$.

\begin{figure}[ht]
	\includegraphics[width=8.5cm]{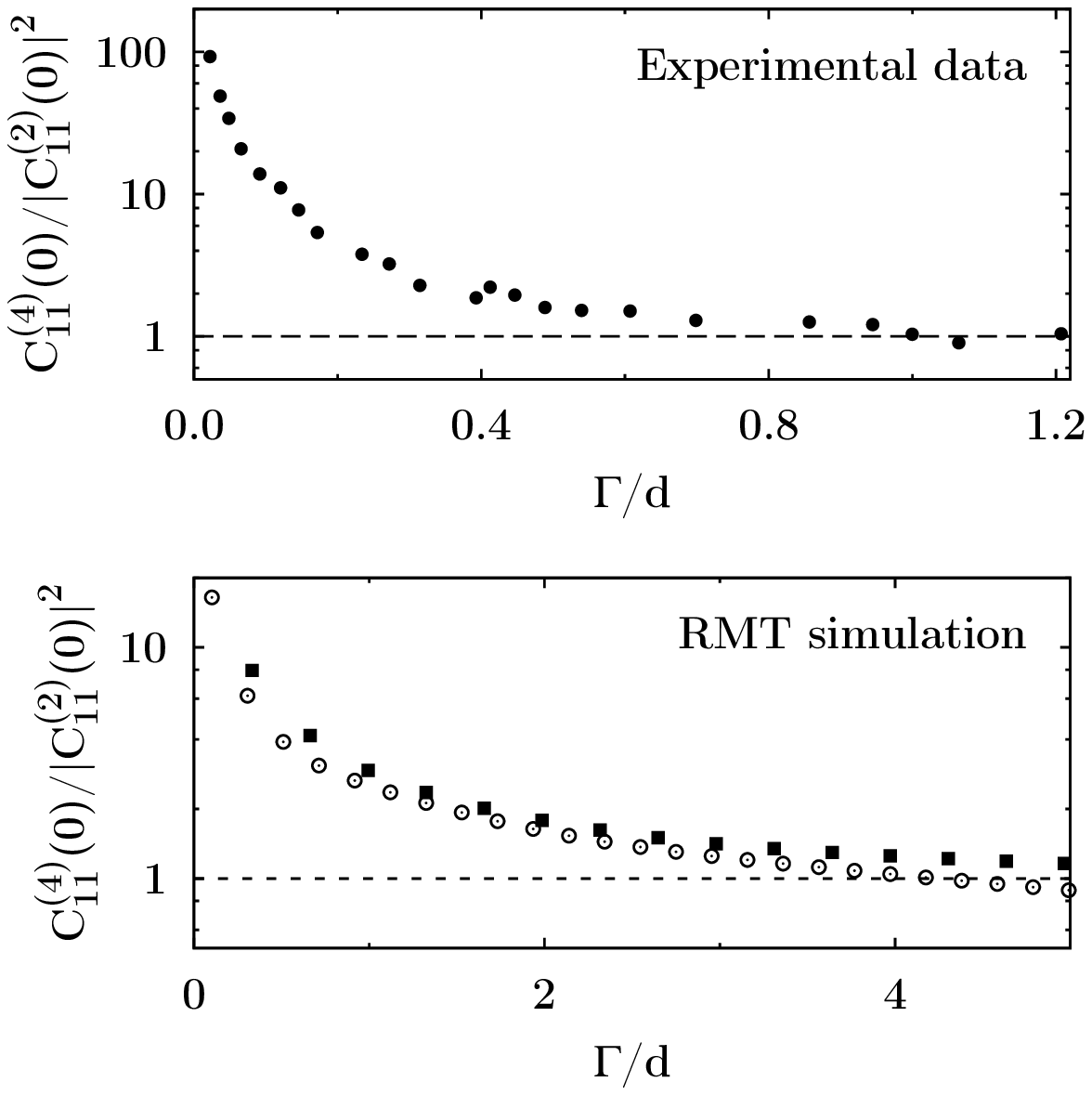}
	\caption{Same as Fig.~\ref{fig1} but for $S^{\rm fl}_{1 1}$.}
	\label{fig4}
\end{figure}

\begin{figure}[ht]
	\includegraphics[width=8.5cm]{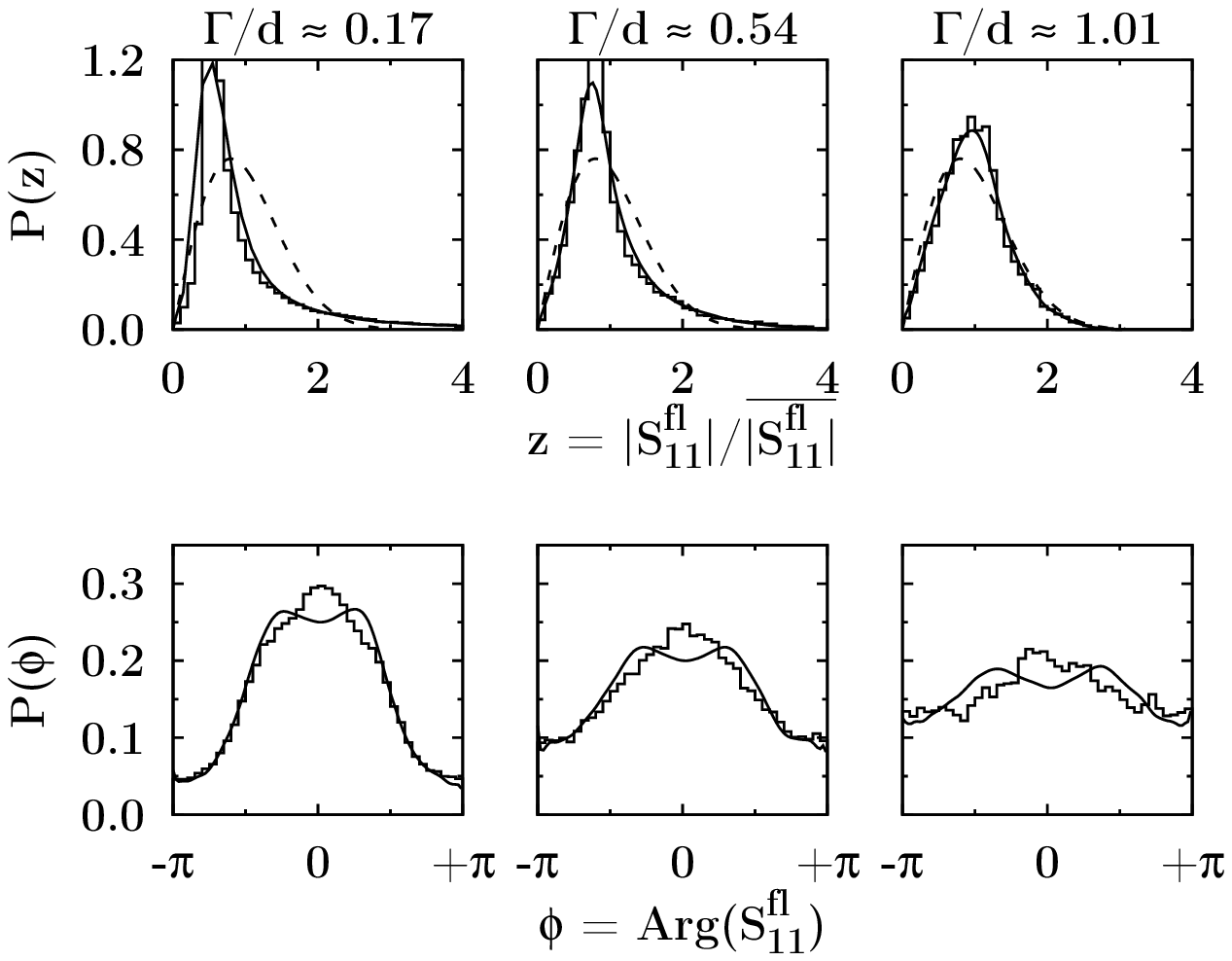}
	\caption{Same as Fig.~\ref{fig2a} but for $S^{\rm fl}_{1 1}$.}
	\label{fig5a}
\end{figure}

\begin{figure}[ht]
	\includegraphics[width=8.5cm]{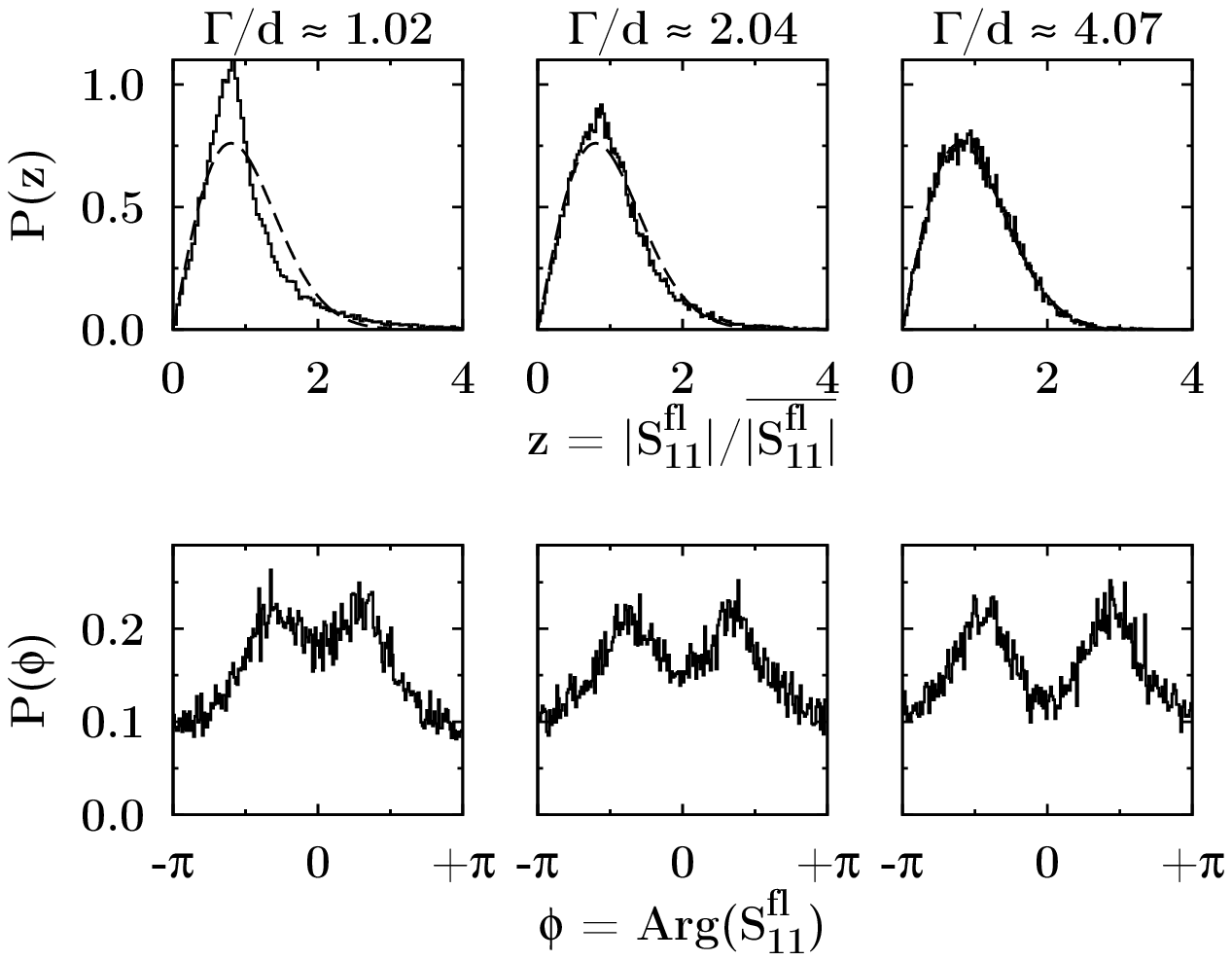}
	\caption{Same as Fig.~\ref{fig2b} but for $S^{\rm fl}_{1 1}$.}
	\label{fig5b}
\end{figure}

\begin{figure}[ht]
	\includegraphics[width=8.5cm]{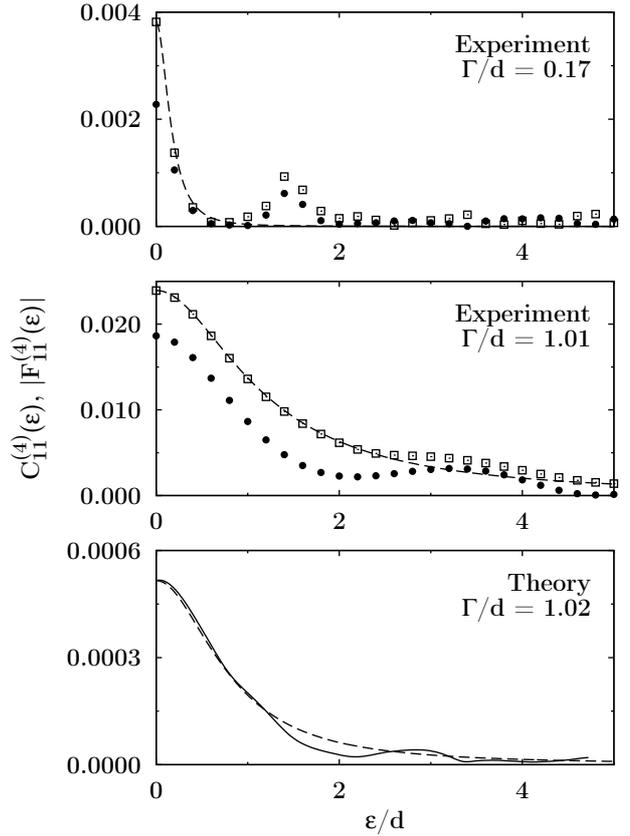}
	\caption{Same as Fig.~\ref{fig3} but for the elastic case.}
	\label{fig6}
\end{figure}

\begin{figure}[ht]
	\includegraphics[width=8.5cm]{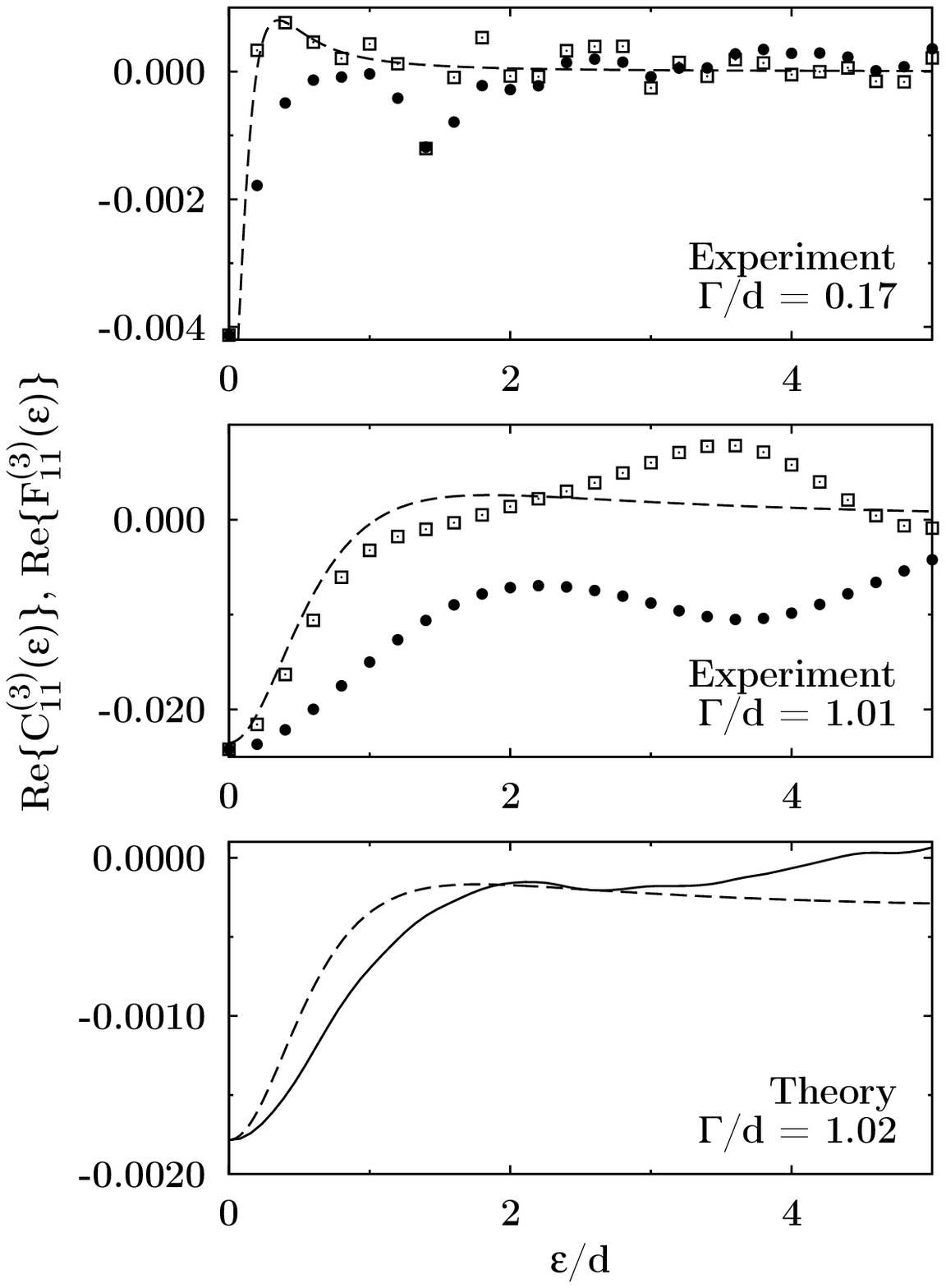}
	\caption{ The dependence on $\ve$ of $\mathfrak{Re}\left(F^{(3)}_{1
	1}(\ve)\right)$ and of $\mathfrak{Re}\left(C^{(3)}_{1 1}(\ve)\right)$
	for three cases with $\Gamma / d$ as indicated in the panels. In the
	upper two panels we show experimental curves for
	$\mathfrak{Re}\left(C^{(3)}_{1 1}(\ve)\right)$ (filled circles) and
	for $\mathfrak{Re}\left(F^{(3)}_{1 1}(\ve)\right)$ (open squares), in
	the lowest panel a numerically generated curve
	$\mathfrak{Re}\left(C^{(3)}_{1 1}(\ve)\right)$ for 32 identical
	channels (solid line). The analytic function
	$\mathfrak{Re}\left(F^{(3)}_{1 1}(\ve)\right)$ is shown as dashed line
	in all three panels. The limited accuracy of our comparison is
	indicated by the difference between the dashed line and the open
	squares in the upper two panels.}
	\label{fig7}
\end{figure}

Concerning $C^{(4)}_{1 1}(0)$, we proceed as in Section~\ref{inel} and
display in Fig.~\ref{fig4} the ratio $C^{(4)}_{1 1}(0) / |C^{(2)}_{1
1}(0)|^2$ in the upper panel for experimental data, in the lower one for
a numerical simulation with $\Lambda=32$ equal transmission
coefficients, as indicated in the figure caption. We reach the same
conclusions as for $a \neq b$: When the transmission coefficients
differ, the ratio is close to unity for $\Gamma \gtrsim d$, when they
are equal for $\Gamma \gtrsim 3d$.
As in the inelastic case, the reason for the failure of Eq.~(\ref{9}) is
the non--Gaussian distribution of $S^{\rm fl}$. This is shown in
Figs.~\ref{fig5a},~\ref{fig5b}. The deviations from a bivariate Gaussian
distribution now occur for larger values of $\Gamma / d$ than in the
inelastic case, in keeping with the expectations formulated at the
beginning of this Section.
In Fig.~\ref{fig6} we show that, as in the elastic case, the rescaled
function $|F^{(4)}_{1 1}(\ve)|$ agrees well with $C^{(4)}_{1 1}(\ve)$
for all values of $\Gamma / d$. Such agreement was likewise found for
the case of $\Lambda = 52$ channels with identical transmission
coefficients. Figure~\ref{fig7} shows that the three-point function
$\mathfrak{Re}\left(C^{(3)}_{1 1}(\ve)\right)$ is approximated quite
well by the (unrenormalized) function $\mathfrak{Re}\left(F^{(3)}_{1
1}(\ve)\right)$ defined in the second of Eqs.~(\ref{Feps}) for the
experimental data with $\Gamma/d=0.17$ (upper panel), where the
resonances are nearly isolated, and for the numerical simulations with
$\Lambda =32$ identical transmission coefficients $T_a=0.2$ (lower
panel). However, considerable deviations are observed for the
experimental data with $\Gamma/d =1.01$ (middle panel). Thus, the
conclusions drawn for the inelastic case apply similarly to the elastic
one only in the regime of isolated resonances and the Ericson regime,
where $\overline{S_{aa}}$ is vanishingly small. To obtain an analytic
expression for the cross-section autocorrelation function we now need to
replace $C^{(4)}_{1 1}(\ve)$ with the rescaled function $|F^{(4)}_{1
1}(\ve)|$ and the function $\mathfrak{Re}\left(C^{(3)}_{1
1}(\ve)\right)$ with $\mathfrak{Re}\left(F^{(3)}_{1 1}(\ve)\right)$.
While the former replacement yields an excellent approximation for all
values of $\Gamma /d$, this is not generally true for the latter in the
intermediate regime of weakly overlapping resonances. Still, for the
numerical simulations with $\Lambda =32$ equal transmission coefficients
$T_a=0.2$, which corresponds to $\Gamma /d=1.02$, the resulting analytic 
expression for the cross-section
autocorrelation function yields a good approximation, because there both,
the values of $\mathfrak{Re}\left(F^{(3)}_{1 1}(\ve)\right)$ and of
$\mathfrak{Re}\left(C^{(3)}_{1 1}(\ve)\right)$, are negligibly small as
compared to ${\cal C}_{a b}(\ve)$, whereas this is not true for the
experimental data with $\Gamma/d=1.01$.

\section{Summary}

We return to the questions raised in the Introduction. We have shown
that Eq.~(\ref{9}) holds essentially only in the Ericson regime and, for
the elastic case, even there only approximately. This is because of the
constraint imposed on the distribution of $S^{\rm fl}$ by unitarity.
Hence, in the inelastic case the cross--section autocorrelation function
can be reliably predicted from Eq.~(\ref{9}) for $\Gamma / d
\gtrsim 1$ for inequivalent channels and for $\Gamma / d \gtrsim 3$ for
identical channels. Even more stringent constraints exist in the elastic
case. We have demonstrated, however, that for the inelastic case for all
values of $\Gamma/d$ and for the elastic case under certain conditions
an excellent approximation to the cross-section autocorrelation function
may be obtained by replacing in Eq.~(\ref{5}) the three-point function
$C^{(3)}_{ab}(\ve )$ by $F^{(3)}_{ab}(\ve )$ and the four-point function
$C^{(4)}_{ab}(\ve )$ by the rescaled function $F^{(4)}_{ab}(\ve )$. With
these replacements, Eq.~(\ref{5}) takes the form
\begin{eqnarray}
	{\cal C}_{a b}(\ve) &\simeq & 2\delta_{a b}\mathfrak{Re}
        \bigg\{\overline{S_{aa}}^2\ C^{(2)}_{a a}(\ve)\nonumber\\ 
        &+&\overline{S_{a a}}\ F^{(3)}_{a a}(\ve) 
        +\overline{S_{a a}}\ F^{(3)}_{a a}(-\ve)\bigg\} \nonumber \\ 
        &+&\frac{C^{(4)}_{a b}(0)}{|F^{(4)}_{a b}(0)|}|F^{(4)}_{a b}(\ve)| 
        \ .
	\label{10}
\end{eqnarray}
The input parameters for the evaluation of Eq.~(\ref{10}) are the average
$S$--matrix elements and the associated transmission coefficients for
all channels and, for the dependence on $\ve$, the average level spacing
$d$ of the scattering system. In terms of these parameters, the
two--point function $C^{(2)}_{a b}(\ve)$ is given in Ref.~\cite{Ver85},
and $C^{(4)}_{a b}(0)$ and the functions $F^{(4)}_{a b}(\ve)$ and
$F^{(3)}_{a b}(\ve)$ in Refs.~\cite{Dav88,Dav89}. For the convenience of
the reader, all relevant formulas are collected in the Appendix in a
form suitable for numerical implementation. For the elastic case,
Eq.~(\ref{10}) provides an excellent approximation for the cross-section
autocorrelation function for all values of $\Gamma /d$ only for $\ve
=0$. For $\ve >0$ Eq.~(\ref{10}) yields a good approximation in the
regime of isolated and strongly overlapping resonances, whereas in the
intermediate regime of weakly overlapping resonances this is true,
if the contribution of $C^{(3)}_{a b}(\ve)$ or equivalently of
$|F^{(3)}_{a b}(\ve)|$ to ${\cal C}_{a b}(\ve)$ is negligible.

\section*{Acknowledgement} We thank T. Friedrich for a discussion at an 
early stage of this work. The research has been supported through the 
SFB~634 by the DFG.

\section*{Appendix}

For the evaluation of ${\cal C}_{a b}(\ve=0)$ in terms of $S^{\rm fl}$
and $\overline{S}$ as given in Eq.~(\ref{5}) we need to compute
$C_{ab}^{(2)}(\ve =0)$, $C_{aa}^{(3)}(\ve =0)= \overline{S_{ab}^{\rm fl
*}(0)\vert S_{ab}^{\rm fl}(0)\vert^2}$ and $C_{ab}^{(4)}(\ve =0)=
\overline{\vert S_{ab}^{\rm fl}(0)\vert^4}-\overline{\vert S_{ab}^{\rm
fl}(0)\vert^2}^2$.  The autocorrelation coefficient $C_{a
b}^{(2)}(\ve=0)$, $\overline{S_{ab}^{\rm fl *}(0)\vert S_{ab}^{\rm
fl}(0)\vert^2}$ and $\overline{\vert S_{ab}^{\rm fl}(0)\vert^4}$ are
given in terms of a threefold integral, c.f. Refs.~\cite{Ver85,Dav88},
\begin{align}
	\label{F}
	& F^{(n)}_{ab}(0) = \\
	& \quad \frac{1}{8} \int_0^\infty{\rm
	d}\lambda_1\int_0^\infty{\rm d}\lambda_2\, \int_0^1 {\rm d}\lambda\,
	\mathcal{J}(\lambda,\lambda_1,\lambda_2) \nonumber \\ 
	& \quad \times \prod_c \frac{1-T_c\, \lambda}{\sqrt{(1+T_c\lambda_1)
	(1+T_c\,\lambda_2)} } \,
	\mathcal{F}^{(n)}_{ab}(\lambda,\lambda_1,\lambda_2) \nonumber
\end{align}
where the integration measure is given as
\begin{align}
	\label{lambdafkt}
	&\mathcal{J}(\lambda,\lambda_1,\lambda_2) = \\
	&\quad \frac{\lambda
	(1-\lambda) \vert\lambda_1-\lambda_2\vert}
	{(\lambda+\lambda_1)^2 (\lambda+\lambda_2)^2
	\sqrt{\lambda_1(1+\lambda_1) \lambda_2
	(1+\lambda_2)}}\, . \nonumber
\end{align}
For $n=2$, i.e. for $\overline{\vert S_{ab}^{\rm fl}(0)\vert^2}$ 
the factor $\mathcal{F}(\lambda,\lambda_1,\lambda_2)$ equals
\begin{align}
	\label{s2}
	& \mathcal{F}^{(2)}_{ab} (\lambda,\lambda_1,\lambda_2) = \\
	&\quad \delta_{ab} \ \vert\overline{S_{aa}}\vert^2 \
		T_a^2 \ \bigg( \frac{\lambda_1}{1+T_a\lambda_1} \nonumber +
		\frac{\lambda_2}{1+T_a\lambda_2}+\frac{2\lambda}{1-T_a\lambda} \bigg)^2 \nonumber \\
	&\quad + (1+\delta_{ab}) \ T_a T_b \  \bigg( \frac{\lambda_1(1+\lambda_1)}
		{(1+T_a\lambda_1)(1+T_b\lambda_1)} \nonumber \\
	&\quad + \frac{\lambda_2(1+\lambda_2)}{(1+T_a\lambda_2)(1+T_b\lambda_2)} +
		\frac{2\lambda(1-\lambda)}{(1-T_a\lambda)(1-T_b\lambda)} \bigg)\, ,
		\nonumber
\end{align}
for $n=3$, i.e. for $\overline{S_{ab}^{\rm fl *}(0)\vert S_{ab}^{\rm
fl}(0)\vert^2}$ it equals
\begin{align}
	\label{s3}
	&\mathcal{F}^{(3)}_{ab} (\lambda,\lambda_1,\lambda_2) = \\
	&\quad -\overline{S}_{aa} \bigg(4{\rm trg} (\mu_a\nu_a) + 2{\rm
		trg}(\mu_a){\rm trg}(\nu_a) \nonumber \\
	&\quad + r_a \left\{{\rm trg} \left(\mu_a^2\right) +
	\frac{1}{2} \left[{\rm trg}(\mu_a)\right]^2 \} \right\} \bigg) \delta_{ab}
	\nonumber
\end{align}
and for $n=4$, i.e. for $\overline{\vert S_{ab}^{\rm fl}(0)\vert^4}$
we have
\begin{align}
	\label{s4}
	& \mathcal{F}^{(4)}_{ab} (\lambda,\lambda_1,\lambda_2) = \\
	&\quad \hspace{1em} \left\{{\rm trg}(\nu_a) + \frac{1}{2}r_a\left[{\rm
		trg}(\mu_a)\right]^2\right\} \nonumber \\
	&\quad \times \left\{{\rm trg}(\nu_b) +
		\frac{1}{2}r_b\left[{\rm trg}(\mu_b)\right]^2\right\} \nonumber \\ 
	&\quad + \delta_{ab} \Bigg( \left[{\rm trg}(\nu_a) \right]^2 + 4{\rm
		trg}\left(\nu_a^2\right) \nonumber \\
	&\quad + r_a^2{\rm trg} \left(\mu_a^2\right) 
	\left\{ {\rm trg} \left(\mu_a^2\right) + \left[{\rm
		trg}(\mu_a) \right]^2\right\} \nonumber\\ 
	&\quad + r_a \Big\{ \left[ {\rm trg}(\mu_a) \right]^2 {\rm
		trg}(\nu_a) \nonumber \\
	&\quad + 8\,{\rm trg}(\mu_a) {\rm trg}(\mu_a\nu_a)
	+ 8\,{\rm trg} \left( \mu_a^2\nu_a \right) \Big\} \Bigg)\, . \nonumber
\end{align}
Here, $r_a=1-T_a$ and $\mu_a$, $\nu_a$ are the matrices
\begin{eqnarray}
	\mu_a &=&
		T_a\lambda_0\left(1+T_a\lambda_0\right)^{-1},\nonumber\\ 
	\nu_a &=&
		T_a^2\lambda_0(1+\lambda_0)\left(1+T_a\lambda_0\right)^{-2}\, ,
\end{eqnarray}
where $\lambda_0$ is the $4\times4$ diagonal matrix with entries
$\lambda_1$, $\lambda_2$, $-\lambda$ and $-\lambda$. For a $4\times4$
matrix $M$ with diagonal elements $m_{ii}$ the graded trace trg is
defined as ${\rm trg}(M)=(m_{11}+m_{22})-(m_{33}+m_{44})$. Analytic
expressions for the functions $F_{ab}^{(3)}(\ve)$ and $F^{(4)}_{a
b}(\ve)$ defined in Eqs.~(\ref{Feps}) are obtained by multiplying
$\mathcal{F}^{(2)}_{ab}(\lambda,\lambda_1,\lambda_2) $ in Eq.~(\ref{s2}) and
$\mathcal{F}^{(3)}_{ab}(\lambda,\lambda_1,\lambda_2) $ in Eq.~(\ref{s3}) and
$\mathcal{F}^{(4)}_{ab}(\lambda,\lambda_1,\lambda_2) $ in Eq.~(\ref{s4})
with the exponential
\begin{equation}
	\exp\left[-i\left(\lambda_1 +\lambda_2+2\lambda\right)\pi\ve/d\right].
	\label{exp}
\end{equation}
The integral Eq.~(\ref{F}) contains several singularities. These can be
handled by proceeding as in Section 5 of Ref.~\cite{Verbaarschot1986}.
We define the variable
\begin{equation}
	p = \lambda_1+\lambda_2+2\lambda \ .
	\label{p}
\end{equation}
Then the exponential in Eq.~(\ref{exp}) turns into
\begin{equation}
	\exp(-ip\pi\ve/d)\, .
	\label{expp}
\end{equation}
We distinguish the two cases $p\leq 2$ and $p>2$. For $p\leq 2$ we
define two further integration variables
\begin{eqnarray}
	s &=& \frac{\sqrt{\lambda_1+\lambda}}{\sqrt{\lambda_1 +
		\lambda_2+2\lambda}}\, ,\\
	t &=& \frac{\sqrt{\lambda_1}}{\sqrt{\lambda_1+\lambda}}\, .
	\label{plt2}
\end{eqnarray}
For the inverse transformation that yields
\begin{eqnarray}
	\lambda   &=& ps^2\left(1-t^2\right)\, ,\\
	\lambda_1 &=& ps^2t^2\, ,\\
	\lambda_2 &=& p\left(1+s^2t^2-2s^2\right)\, .
	\label{lamplt2}
\end{eqnarray}
The Jacobian for this transformation is equal to $4p^2s^3t$. The 
threefold integral in Eq.~(\ref{F}) becomes
\begin{align}
	\label{F1}
	& F^{(n)}_{ab}(\ve)_{(p\leq 2)} = \\
	& \int_0^2{\rm d}p\, \int_0^{1/\sqrt{2}}{\rm d}s\, \int_0^1 {\rm
	d}t\,\frac{e^{-ip\pi\ve/d}}{p\left(1-s^2\right)^2
	\sqrt{\left(1+ps^2t^2\right)}} \nonumber \\
	& \quad \times \,\frac{\left[1-ps^2\left(1-t^2\right)\right]
	\left(1-2s^2\right)\left(1-t^2\right)} {\sqrt{\left(1+s^2t^2-2s^2\right)
	\left[1+p\left(1+s^2t^2-2s^2\right)\right]}} \nonumber \\ 
	& \quad \times \, \prod_c\frac{1-T_c\, \lambda}{\sqrt{(1+T_c\lambda_1)
	(1+T_c\,\lambda_2)} } \, \mathcal{F}^{(n)}_{ab}(\lambda,\lambda_1,\lambda_2)
	\, . \nonumber
\end{align}

For $p>2$ we define additional integration variables 
\begin{eqnarray}
	s&=&\sqrt{\lambda_1+\lambda}\, ,\\
	t&=&\frac{\sqrt{\lambda_1}}{\sqrt{\lambda_1+\lambda}}\, .
	\label{pgt2}
\end{eqnarray}
For the inverse transformation that yields
\begin{eqnarray}
	\lambda   &=& s^2\left(1-t^2\right)\, ,\\
	\lambda_1 &=& s^2t^2\, ,\\
	\lambda_2 &=& p+s^2t^2-2s^2\, .
	\label{lampgt2}
\end{eqnarray}
The Jacobian equals $4s^3t$ and the threefold integral in Eq.~(\ref{F})
becomes
\begin{align}
	& F^{(n)}_{ab}(\ve)_{(p>2)} =\\ 
	& \int_2^\infty{\rm d}p\, \int_0^{\sqrt{p/2}}{\rm d}s \,
	\int_{\sqrt{1-s^{-2}}\theta(s-1)}^1 {\rm d}t\,
	\frac{e^{-ip\pi\ve/d}}{\left(p-s^2\right)^2} \nonumber \\
	& \quad \times \, \frac{\left[1-s^2\left(1-t^2\right)\right]
	\left(p-2s^2\right)\left(1-t^2\right)}
	{\sqrt{\left(p+s^2t^2-2s^2\right)\left(1+s^2t^2\right)
	\left(1+p+s^2t^2-2s^2\right)}} \nonumber \\
	& \quad \times \, \prod_c\frac{1-T_c\, \lambda}
	{\sqrt{(1+T_c\lambda_1)(1+T_c\,\lambda_2)} } \,
	\mathcal{F}^{(n)}_{ab}(\lambda,\lambda_1,\lambda_2) \nonumber
	\label{F2}
\end{align}
where the $\theta$-function is defined by
\begin{eqnarray}
	\theta (x)&=&0\, \, {\rm for}\, \, x\leq 0\, , \\
	\theta (x)&=&1\, \, {\rm for}\, \, x> 0\, .
\end{eqnarray}

\end{document}